\documentclass[10pt,letterpaper]{article}
\usepackage{opex3}

\usepackage{graphicx}
\usepackage{bm}
\usepackage{amssymb}
\usepackage{amsmath}
\usepackage{float}
\usepackage{pstricks} 
\usepackage{color}
\usepackage{bbold}
\usepackage{xr} 

\graphicspath{{/}}

\newcommand{\dg}{\dagger}
\def\vec#1{\boldsymbol{#1}}

\begin{document}

\title{On-chip low loss heralded source of pure single photons}

\author{Justin B. Spring$^{1,*}$, Patrick S. Salter$^2$, Benjamin J. Metcalf$^1$, Peter C. Humphreys$^1$, Merritt Moore$^1$, Nicholas Thomas-Peter$^1$, Marco Barbieri$^1$, Xian-Min Jin$^{1,3}$, Nathan K. Langford$^{1,4}$, W. Steven Kolthammer$^1$, Martin J. Booth$^{2,5}$, and Ian A. Walmsley$^{1,*}$}

\address{$^1$Clarendon Laboratory, University of Oxford, Parks Road, Oxford OX1 3PU, United Kingdom \\ $^2$Department of Engineering Science, University of Oxford, Parks Road, \\ Oxford OX1 3PJ, United Kingdom \\  $^3$ Department of Physics, Shanghai Jiao Tong University \\ Shanghai 200240, People's Republic of China \\  $^4$ Department of Physics, Royal Holloway, University of London, \\ London TW20 0EX, United Kingdom \\ $^5$ Centre for Neural Circuits and Behaviour, University of Oxford, Mansfield Road, \\ Oxford OX1 3SR, United Kingdom \\ *Corresponding authors: j.spring1@physics.ox.ac.uk, i.walmsley1@physics.ox.ac.uk}




\begin{abstract}
A key obstacle to the experimental realization of many photonic quantum-enhanced technologies is the lack of low-loss sources of single photons in pure quantum states. We demonstrate a promising solution: generation of heralded single photons in a silica photonic chip by spontaneous four-wave mixing. A heralding efficiency of $40\%$, corresponding to a preparation efficiency of $80\%$ accounting for detector performance, is achieved due to efficient coupling of the low-loss source to optical fibers. A single photon purity of $0.86$ is measured from the source number statistics without filtering, and confirmed by direct measurement of the joint spectral intensity. We calculate that similar high-heralded-purity output can be obtained from visible to telecom spectral regions using this approach.  On-chip silica sources can have immediate application in a wide range of single-photon quantum optics applications which employ silica photonics.
\end{abstract}


\bibliographystyle{osajnl}

%
%

\section{Introduction}
Photonic quantum-enhanced technologies aim to employ nonclassical states of light to surpass classical performance limits in diverse fields including computation~\cite{Knill2001,Spring2013}, metrology~\cite{Giovannetti2004}, and communication~\cite{Briegel1998}.  An impediment to faster progress is the quality of available single photon sources.   Building the low-loss sources of high purity single photons necessary for quantum-enhanced performance has proven challenging~\cite{Eisaman2011,Varnava2008,Datta2011,Jennewein2011,Lucamarini2012}.

Heralding spontaneous emission from a nonlinear-optical material has to date been the most common method of generating single photons~\cite{Bouwmeester1998,Halder2009,Soller2011,Metcalf2013}.  Nonlinear processes such as spontaneous parametric down-conversion (SPDC) or spontaneous four-wave mixing (SFWM) can be used to create pairs of photons.  Due to this pairwise emission, detection of one photon, the heralding photon, indicates the creation of its partner, the heralded single photon.  A key source metric is the preparation efficiency $\eta_P$, the conditional probability that a heralded photon is delivered to its application given detection of the heralding photon.  For example, the rate at which multiphoton states are generated from single-photon sources scales exponentially with $\eta_P$, even with multiplexing strategies which ideally achieve near-deterministic emission~\cite{Christ2012}.  In numerous applications, $\eta_P$ is a crucial parameter for a quantum method to demonstrate true advantage over a classical approach~\cite{Varnava2008,Datta2011,Jennewein2011,Lucamarini2012}.

Four-wave mixing in a silica waveguide is a promising route to achieving exceptionally high $\eta_P$~\cite{Soller2011,SmithBJ2009ppg}.  In heralded photon sources, reduction in $\eta_P$ results from loss of the heralded photon, due to scattering or imperfect mode-matching at interfaces.  Waveguides in commonly available silica glasses minimize these effects due to their exceedingly low optical loss and excellent mode-matching to optical fiber, a ubiquitous component in quantum photonics.  While silica's relatively small $\chi^{(3)}$ nonlinearity affects the emitted photon flux, in many applications it is loss that is fundamental to demonstrating quantum enhancement, not flux.  In fact, for heralded single photon sources one must deliberately keep emission rates low to avoid unwanted heralding of more than one photon.  The transverse confinement of the waveguide allows these desired photon production rates to be reached at readily available pump powers.

Typical quantum applications require heralded photons in pure quantum states in addition to high source $\eta_P$.  In general, however, the heralded pair source generates mixed quantum states.  Energy and momentum conservation can lead to entanglement between multiple spatial and spectral modes of the emitted photon pairs~\cite{GriceWP2001efs,U'Ren2005}.  If the heralding photon is detected but its mode is not resolved, then the heralded photon is left in a mixed quantum state of all possible modes.  One approach to achieve a high-purity heralded photon is to use spectral and spatial filters on the heralding field which ensures the detector responds to only a single spatio-temporal mode.  Such filters reduce the rate at which heralded photons are emitted.  On-chip SPDC sources are capable of high photon flux~\cite{MosleyPJ2009dms,Zhong2009,Karpinski2012,EcksteinA2011hes} that allows acceptable count rates even when filters are used to achieve good photon purity.  For SFWM in silica, on the other hand, the reduced count rate from this approach could lead to unacceptably long data collection times.

An alternate approach we employ here is to engineer the source to emit photons into a single pair of modes.  As a consequence, the heralding and heralded photon are not entangled.  The resulting factorable output allows heralded photons of high purity without filtering.  Guided-wave photonics allows the precise control of optical modes needed to construct such a source.  Previous silica sources have demonstrated this strategy using optical fibers.  In these structures, however, fabrication inhomogeneity~\cite{Halder2009,Cohen2009} or sensitivity to the local environment~\cite{Soller2011,Zhang1993,Zou2009,SollerThesis} prevented a robust scalable solution.  While on-chip SFWM sources have been demonstrated in chalcogenide glass~\cite{Xiong2011a} and silicon~\cite{Davanco2012}, these devices exhibited relatively high loss and did not pursue the factorable source design strategy described here.

Here we report the first photon source on a silica photonic chip.  Our SFWM source generates heralded single photons of purity $P=0.86$ and $\eta_P=80\%$ without any filtering.  Heralded single photons are emitted at a rate of $3.1\cdot10^5$ photons per second with a pump field power of $150$ mW.  We use a birefringent waveguide to phasematch SFWM at frequencies where the red-detuned photon lies far beyond silica's Raman gain peak.  This minimizes spontaneous Raman scattering~\cite{Lin2007,Soller2011,SmithBJ2009ppg}, which is the principal noise source in many silica sources~\cite{Fiorentino2002,Inoue2004}.  A solid-clad silica waveguide provides a convenient and robust platform for our source which we foresee finding immediate application in integrated quantum optics experiments that frequently rely on similar integrated silica architectures~\cite{Spring2013,Metcalf2013,Marshall2009,Shadbolt2011,Crespi2011,Owens2011,Crespi2013}.

\section{Experimental overview}
Our source uses a $4$ cm long waveguide fabricated by femtosecond laser writing in an undoped silica chip (Lithosil Q1)~\cite{Davis1996,DellaValle2009}.  We use adaptive optics to shape the writing beam~\cite{Salter2012} and produce an elliptical transverse mode yielding a birefringence of $\Delta n=10^{-4}$.  By measuring the insertion loss and imaging the spatial mode, we determine the propagation loss in our waveguide to be less than $0.4$ dB/cm~\cite{Salter2012}.

\begin{figure}[htbp]
\centering\includegraphics[width=13cm]{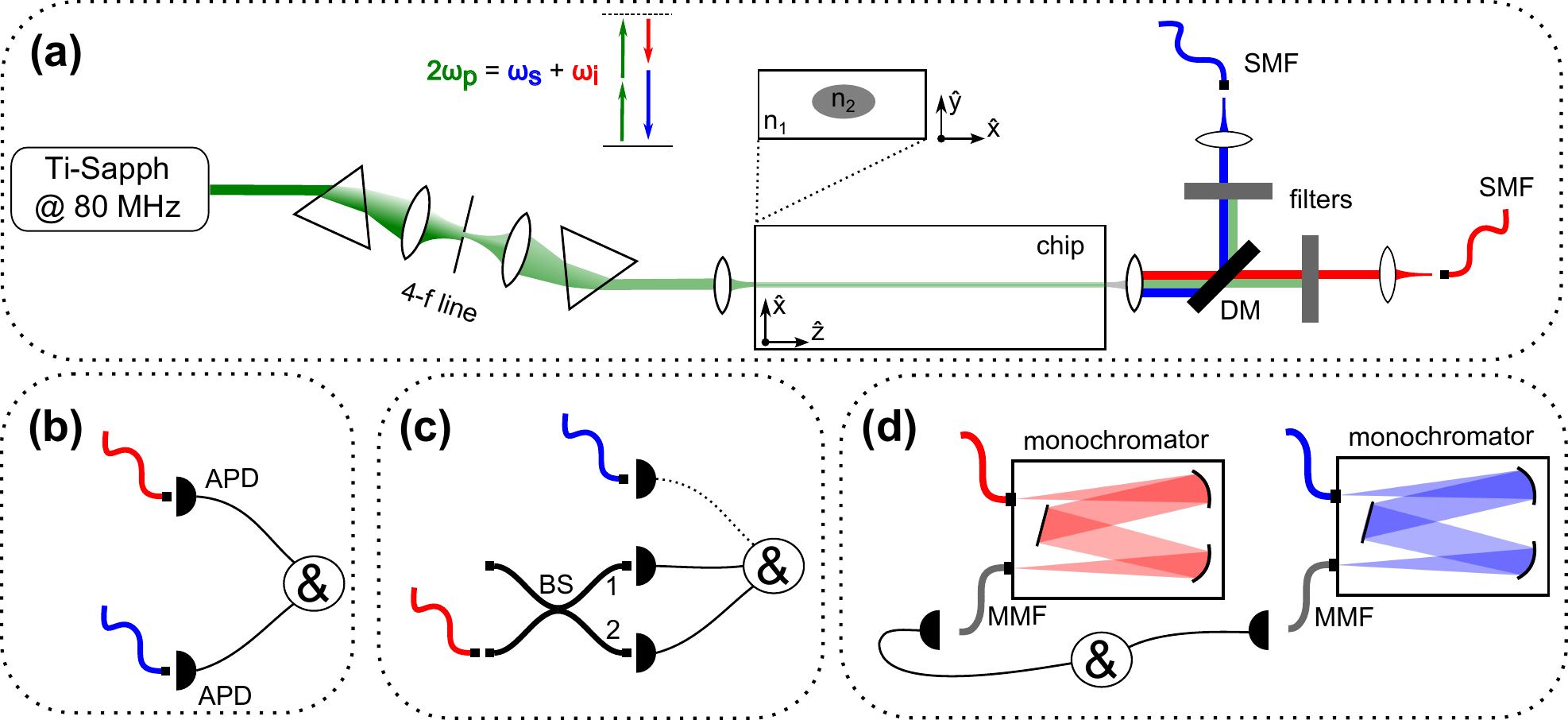}
\caption{The experimental layout consists of (a) state preparation and (b-d) characterization.  (a) A 4-f line with an adjustable slit is used to control the bandwidth of the pump ($\Delta\lambda_p$).  After the chip, the pump is filtered out while the signal and idler fields are separated by a dichroic mirror (DM) and coupled into separate single-mode fibers (SMFs). (b) Source count rate and cross-correlation $g^{(2)}_{si}(0)$ are measured with the SMFs coupled directly into avalanche photodiodes (APDs).  (c) The autocorrelation $g^{(2)}_{ii}(0)$ is measured by connecting the idler fiber to a fiber beam splitter (BS) with a reflectivity of $50\%$ and ignoring the signal field, while the signal field is used as a herald when determining $g^{(2)}_H(0)$.  We measure $g^{(2)}_{ss}(0)$ by inserting the signal fiber into the BS and ignoring the idler field (not shown).  (d) To measure joint spectral intensities, the source SMFs are connected to separate monochromators with multi-mode fibers (MMFs) at the output performing a raster scan over the joint spectral range while APDs count in coincidence.}
\label{fig:setup}
\end{figure}

A single pump field with central wavelength $\lambda_p=729$ nm is generated by an $80$ MHz Ti-Sapphire oscillator whose spectral bandwidth $\Delta\lambda_p$ is adjusted with a mechanical slit in a 4-f line, as shown in Fig.~\ref{fig:setup}(a).  We describe the high (low) frequency photon in the emitted pair as the signal (idler).  A dichroic beam splitter (Semrock FF740) separates these signal and idler photons which have central wavelengths of $\lambda_s=676$ and $\lambda_i=790$ nm.  A combination of spectral (Semrock 684/24 and 800/12) and polarization filters extinguish the pump field before the signal and idler modes are coupled into single-mode fibers.  Silicon avalanche photodiodes (APDs) (Perkin Elmer SPCM-AQ4C) and an FPGA-based coincidence counter (Xilinx SP-605) are used to measure marginal and joint photon statistics as shown in Fig.~\ref{fig:setup}(b)-(d) and discussed below.

\section{Nonclassical emission and heralding of single photons}
We first confirm nonclassical operation of our source by measuring the cross- and auto-correlations of the signal and idler modes with the setup shown in Fig.~\ref{fig:setup}(b)-(c).  Classical fields must satisfy the Cauchy-Schwarz inequality $(g^{(2)}_{si}(0))^2 \leq g^{(2)}_{ss}(0) \cdot g^{(2)}_{ii}(0)$ where $g^{(2)}_{xy}(\tau)$ is the second order coherence between modes $x$ and $y$ at relative time delay $\tau$~\cite{Loudon2000}. For our pulsed source, we calculate
\begin{equation}
g^{(2)}_{xy}(0) = \frac{N_{xy}N_{p}}{N_{x}N_{y}}
\label{eq:g2}
\end{equation}
where $N_{x}$ ($N_{xy}$) correspond to single (temporally coincident) detection events on mode $x$ ($x$ and $y$).  The number of trials is given by $N_{p}$, the number of pulses from the Ti-sapphire pump.  Autocorrelations are described by $g^{(2)}_{x_1x_2}(0)$ where $x_{1,2}$ refer to the two output ports of a fiber beamsplitter with a reflectivity of $50\%$ as shown in Fig.~\ref{fig:setup}(c).

\begin{figure}[htbp]
\centering\includegraphics[width=10cm]{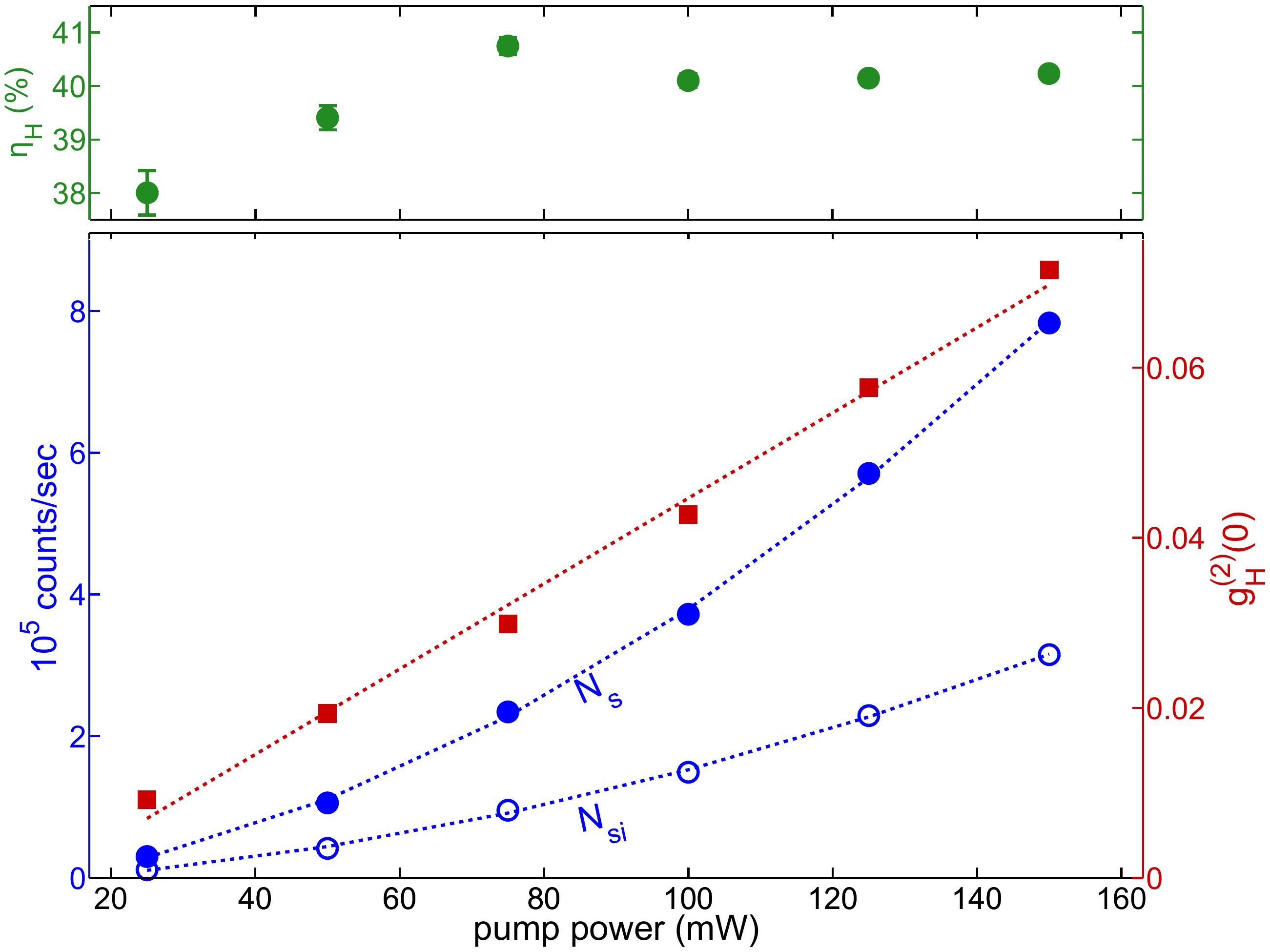}
\caption{The count rate for the signal mode alone ($N_s$) and in temporal coincidence with the idler mode ($N_{si}$) is shown as a function of the pump power (blue circles).  The heralding efficiency, which includes APD detector losses, is calculated according to $\eta_H = N_{si}/N_s$ (green).   At low pump powers, $\eta_H$ decreases due to the heightened importance of detector dark counts, which cause false herald events.  The heralded autocorrelation of the idler photon with the signal field as herald, $g^{(2)}_H(0)$, increases linearly due to spontaneous Raman scattering (red squares).  Quadratic (blue) and linear (red) fits to the data are shown with dotted lines. Error bars are smaller than the markers for the count data and $g^{(2)}_H(0)$.}
\label{fig:cntsHerG2}
\end{figure}

With the pump bandwidth $\Delta\lambda_p = 3.1$ nm and $100$ mW average power, we measure $g^{(2)}_{si}(0)=73.5\pm1.1$, $g^{(2)}_{ss}(0) = 1.82\pm0.03$, and $g^{(2)}_{ii}(0) = 1.26\pm0.02$ with no background subtraction.  The Cauchy-Schwarz inequality is violated here by 49 standard deviations which demonstrates a nonclassical correlation between signal and idler modes in the photon number basis.  

It is this correlation in photon number that makes such sources suitable for heralding single photons.  Using the signal photon as the heralding photon, we measure the heralding efficiency $\eta_H=N_{si}/N_s$.  We calculate $\eta_H>40\%$ for pump powers of $75$ to $150$ mW as seen in Fig.~\ref{fig:cntsHerG2}.  Since $\eta_H$ is limited by the detector efficiency, which is not our concern here, we estimate $\eta_P$, which corresponds to the probability that the idler photon arrives at its detector given detection of the heralding signal photon.  Using the manufacturer's specified detector efficiency, we estimate $\eta_P=80\%$.  The remaining inefficiency is primarily due to loss from coupling to single mode fiber, reflection at interfaces, and scattering.  Neither the chip nor the fibers were AR-coated.    

We quantify the suppression of undesired higher order photon emission via the heralded second-order correlation at zero relative time delay
\begin{equation}
g^{(2)}_H(0) = \frac{N_{i_1i_2s}N_s}{N_{i_1s}N_{i_2s}}
\label{eq:herG2}
\end{equation}
for which all counts are conditional on detecting a heralding signal photon as shown in Fig.~\ref{fig:setup}(c).  An ideal single-photon source would give $g^{(2)}_H(0) = 0$. We measure $g^{(2)}_H(0) = 0.0092\pm0.0004$ for a pump power of $25$ mW.  Spontaneous Raman scattering is the primary noise source, which explains the linear increase in $g^{(2)}_H(0)$ as the source is pumped harder.  Our low $g^{(2)}_H(0)$ values are comparable to other SFWM sources~\cite{SmithBJ2009ppg,SollerThesis} that take advantage of birefringent phase matching~\cite{Lin2007}, which allows significant suppression of this Raman noise which has inhibited other SFWM sources~\cite{Fiorentino2002,Inoue2004}.  

\section{Controlling the heralded state purity}
The pairwise emission that allows heralding of single photons is not sufficient for many applications; these photons must also be in pure states.  Undesired correlations between the signal and idler fields that are not resolved by the heralding detector instead result in heralding of mixed states.  Such correlations imply multimode emission in the frequency, polarization, or spatial degrees of freedom.  Phasematching constraints generally force the signal and idler fields to be emitted into single polarizations.  Furthermore, our waveguide constrains each of the three fields (pump, signal, idler) to a single spatial mode.  This is more readily achieved for a SFWM source, in contrast to  SPDC, due to the reduced disparity in field frequencies.

We focus on the remaining possibility that correlations are generated in the spectral degree of freedom.  The effective SFWM Hamiltonian can thus be approximated as~\cite{GriceWP2001efs}
\begin{equation}
\hat{H}_{\text{SFWM}}=\zeta\iint d\omega_s d\omega_i f(\omega_s,\omega_i)\hat{a}^\dg(\omega_s)\hat{b}^\dg(\omega_i) + \text{H.c.}
\label{eq:Hamiltonian}
\end{equation}
where the joint spectral amplitude $f(\omega_s,\omega_i){=}\int d\omega'\alpha(\omega')\alpha(\omega_s{+}\omega_i{-}\omega')\phi(\omega_s,\omega_i)$ is a function of the pump envelope $\alpha(\omega)$ and phasematching function $\phi(\omega_s,\omega_i)$, while $\zeta$ depends on both the pump intensity and magnitude of the $\chi^{(3)}$ nonlinearity.  We approximate $\alpha(\omega)$ as a Gaussian function and specify $\Delta\lambda_p$ as the full-width at half maximum of $|\alpha(\omega)|^2$.  The phasematching function results from integrating over the length of the guide $\phi(\omega_s,\omega_i)=\int_0^L e^{i\Delta k z} dz\propto e^{i\Delta k L} \textrm{sinc}(\Delta kL/2)$, where the wavevector mismatch is $\Delta k = 2k_p-k_s-k_i$.  We neglect the phase mismatch arising from the pump pulse intensity, since this is small for our source.  In the normal dispersion regime, SFWM is phase-matched ($\Delta k=0$) when the pump field is polarized along the slow axis and both signal and idler fields are polarized along the fast axis of the birefringent waveguide.

The spectral entanglement generated by the Hamiltonian in Eq.~\ref{eq:Hamiltonian} can be viewed as a consequence of energy and momentum conservation.  To quantify these correlations, we rewrite the Hamiltonian in terms of a minimal set of broadband modes via the Schmidt decomposition~\cite{Law2000}
\begin{equation}
\hat{H}_{\text{SFWM}} {=} \sum_{m=1}^N c_m \hat{A}^\dg_m(\omega_s)\hat{B}^\dg_m(\omega_i) + \text{H.c.}
\label{eq:Schmidt}
\end{equation}
where $\hat{A}_m^\dg(\omega_s)=\int d\omega_s \xi_m(\omega_s)\hat{a}^\dg(\omega_s)$ and $\hat{B}_m^\dg(\omega_i)=\int d\omega_i \psi_m(\omega_i)\hat{a}^\dg(\omega_i)$.  The set of functions $\{\xi_m\}$, $\{\psi_m\}$ are called Schmidt modes, which define an orthonormal basis for the signal and idler Hilbert spaces respectively.  This decomposition shows the evolution due to SFWM is equivalent to an ensemble of two-mode squeezing operators $e^{i \hat{H}_{\text{SFWM}}} {=} \hat{S}_{A_1,B_1}\otimes\hat{S}_{A_2,B_2}\otimes\text{...}$ where $\hat{S}_{A_m,B_m}$ is a two-mode squeezing operator on modes $A_m$ and $B_m$~\cite{EcksteinA2011hes}.  

In general, the Schmidt decomposition in Eq.~\ref{eq:Schmidt} includes significant contributions from multiple modes so that $N>1$.  As previously discussed, a detector that cannot resolve these different frequency modes leads to heralding of mixed quantum states.  To restore high purity, one can employ spectral filters to remove higher modes but at the cost of reduced count rate.  In contrast, the approach we adopt is to design the source to emit only in a single pair of Schmidt modes~\cite{GriceWP2001efs,U'Ren2005,Soller2011,MosleyPJ2008hgu}.  This allows heralding of high purity states without filtering.  We use the singular value decomposition to numerically perform the Schmidt decomposition in Eq.~\ref{eq:Schmidt}.  This allows one to predict the heralded photon purity given source parameters $\alpha(\omega)$ and $\phi(\omega_s,\omega_i)$~\cite{MosleyPJ2008hgu}.

\begin{figure}[htbp]
\centering\includegraphics[width=13cm]{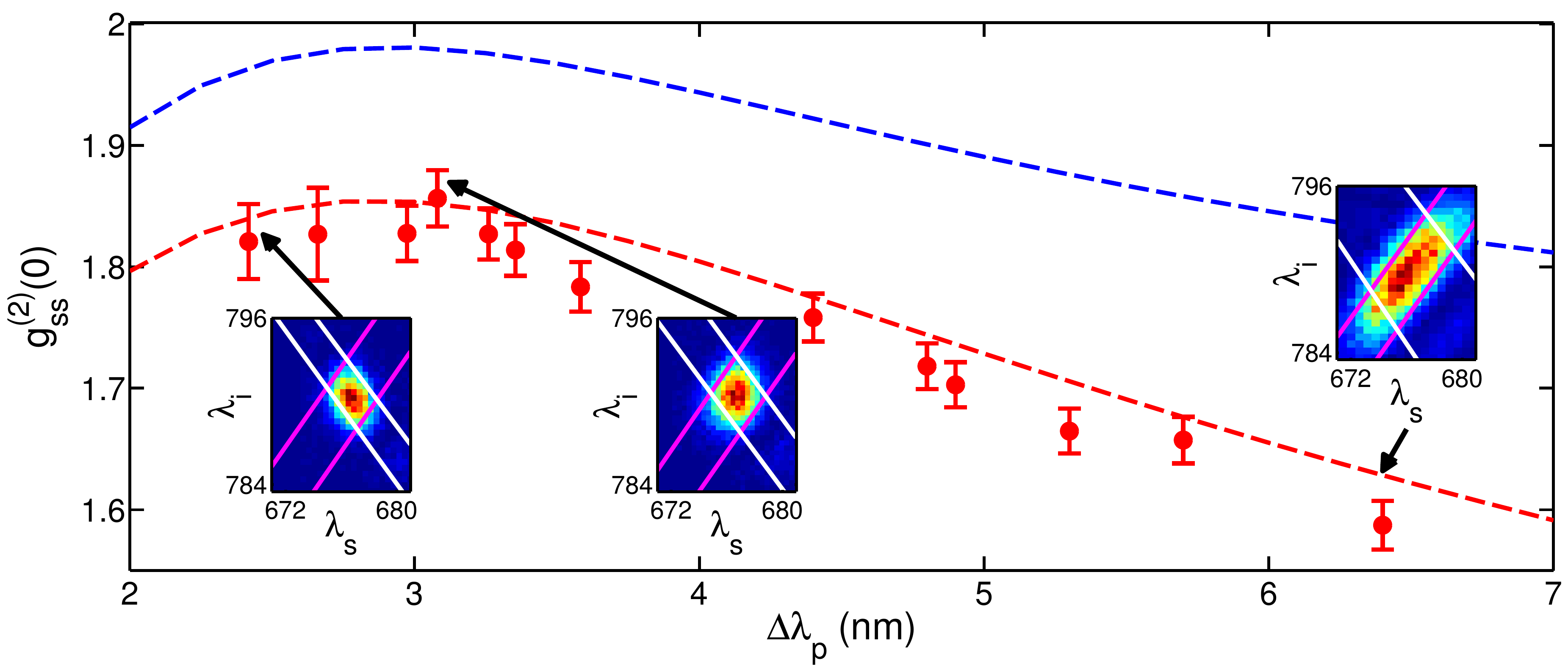}	
\caption{Unheralded $g^{(2)}_{ss}$ of the signal field shows control over the amount of spectral entanglement between the signal and idler fields as $\Delta\lambda_p$ is adjusted (red).  Theoretical curve (dashed) and data (points) are shown.  Filtering out the peripheral lobes in the joint spectral intensity with a $4.5$ nm filter on the signal field is calculated to give $g^{(2)}_{ss} = 1.98$ at $\Delta\lambda_p = 3$ nm (blue).  Insets: joint spectral intensity (JSI) measurements demonstrate spectral entanglement control. The FWHM of the pump envelope $|\alpha|^4$ (white) and phase-matching function $|\phi|^2$ (purple) accurately predict JSI orientation.}
\label{fig:g2jntSpec}
\end{figure}

To investigate the frequency-mode structure of our source, we measure the marginal photon number distribution.  An ideal single-mode source would exhibit a thermal distribution, while a highly multi-mode emitter gives Poissonian statistics~\cite{Mauerer2009}. In our source, the transition from single-mode to increasingly multi-mode behavior can be readily adjusted via the pump bandwidth $\Delta\lambda_p$~\cite{SmithBJ2009ppg}.  To demonstrate this, we measure the autocorrelation $g^{(2)}_{ss}(0)$ as $\Delta\lambda_p$ is varied, as shown in Fig.~\ref{fig:g2jntSpec}.  For $\Delta\lambda_p = 3.1$ nm we measure our optimal $g^{(2)}_{ss}(0) = 1.86\pm0.02$ which is close to the ideal thermal result $g^{(2)}(0)=2$.  Only APD dark counts are subtracted from the data used to calculate $g^{(2)}_{ss}(0)$ in Fig.~\ref{fig:g2jntSpec}.   One can relate the number of excited Schmidt modes to these statistics using $g^{(2)}_{ss}(0) = 1+1/\sum_m |c_m|^4=1+P$ where $P$ is the heralded purity.  Thus, we have demonstrated $P = 0.86$ without any filtering of the signal and idler modes.  To our knowledge, no previous on-chip source has simultaneously demonstrated purities and efficiencies as high as $P = 0.86$ and $\eta_p=80\%$.  

The near single-mode emission of our source is further supported by joint spectral intensity measurement.  A spectrally uncorrelated state has a factorable joint spectral amplitude, and thus a factorable joint spectral intensity.   In the middle inset of Fig.~\ref{fig:g2jntSpec}, we find that the major and minor axes of the central ellipse are parallel to the $\lambda_{i,s}$ axes, which shows a factorable intensity $I(\lambda_i,\lambda_s)=I_i(\lambda_i)I_s(\lambda_s)$.  As $\Delta\lambda_p$ is adjusted away from this optimal value, the joint spectral intensities in the left and right insets of Fig.~\ref{fig:g2jntSpec} indicate entanglement with an increasingly tilted central lobe.  These spectral correlations are in agreement with the measured decrease in $g^{(2)}_{ss}(0)$. 

At the optimal $\Delta\lambda_p$, the slight deviation from ideal thermal statistics, and corresponding reduction in heralded state purity, is principally due to peripheral lobes in the phasematching function.  These arise from the hard-edge boundaries of the waveguide and are faintly observed in joint spectral intensity plots.  A $4.5$ nm filter on $\lambda_s$ would suppress the small lobes, which would yield $g^{(2)}_{ss}(0)=1.98$, and corresponding $P=0.98$, while still passing $90\%$ of photons.  Such a filter leaves $\eta_P$ unchanged, as it would only be applied to the heralding signal photon.

\begin{figure}[htbp]
\centering\includegraphics[width=13cm]{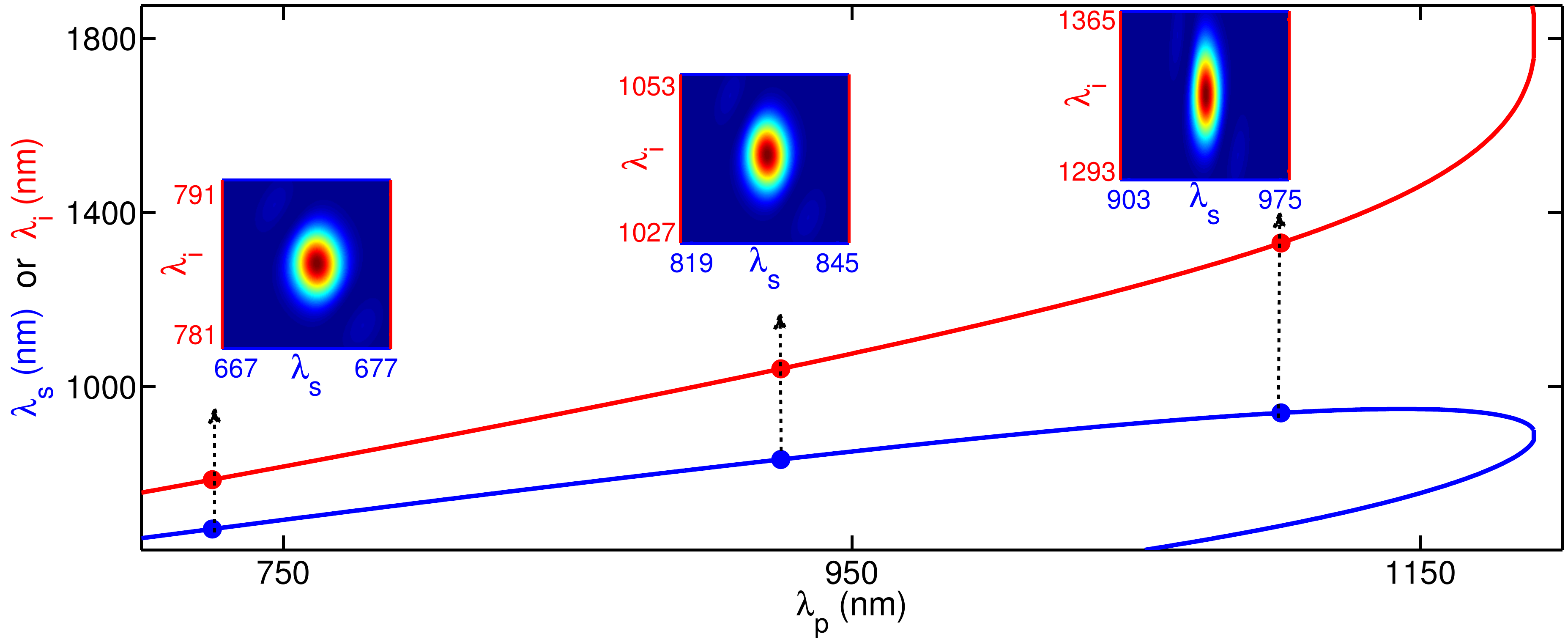}	
\caption{Theoretical phasematching curves show the signal (blue) and idler (red) wavelengths that satisfy the condition $\Delta \vec{k} = 0$ for a range of pump wavelengths.  Insets show predicted SFWM joint spectral intensities.  A factorable output suitable for heralded production of pure photons can be produced over a wide spectral range, from visible to telecom wavelengths, by adjusting $\lambda_p$ and $\Delta\lambda_p$ (left to right: $\Delta \lambda_p = 3.1$,~$7.5$,~$20$ nm).  These calculations assume L = $4$ cm and $\Delta n=10^{-4}$.}
\label{fig:phasematch}
\end{figure}

Our source offers large spectral tunability of the signal and idler fields.  Fig.~\ref{fig:phasematch} shows that SFWM is phase-matched over a wide range of pump wavelengths.  The inset figures illustrate that over this entire range $\Delta\lambda_p$ can be adjusted to achieve factorable output.  For larger $\lambda_p$ the group velocities of the pump and idler photons become comparable, which results in a large spectral bandwidth for the idler.  This relatively simple route to broadband emission could be useful, for example in quantum optical coherence tomography~\cite{Nasr2008}.  In contrast, many applications, such as quantum memories, instead require narrowband emission.  A theory for cavity-enhanced SFWM has been developed that allows emission down to MHz bandwidths~\cite{Garay-Palmett2013}.  Cavities can be implemented using the refined Bragg grating technology available in silica~\cite{Marshall2006,Lepert2011}.

\section{Birefringence homogeneity}
Our analysis so far assumes the waveguide, and thus the wavevectors $\vec{k_{p,s,i}}$, are constant throughout the source.  Imperfections in this regard can diminish the purity of heralded photons.  Due to our operation of a solid-clad guide far from its zero dispersion wavelength along with the excellent material uniformity of commercial silica~\cite{Stolen1981,SmithBJ2009ppg}, the dominant source of inhomogeneity is the birefringence.  The spectral output is extremely sensitive to variation in the birefringence since the phase-matched wavelengths depend critically on this parameter~\cite{Soller2011}.

\begin{figure}[htbp]
\centering\includegraphics[width=12cm]{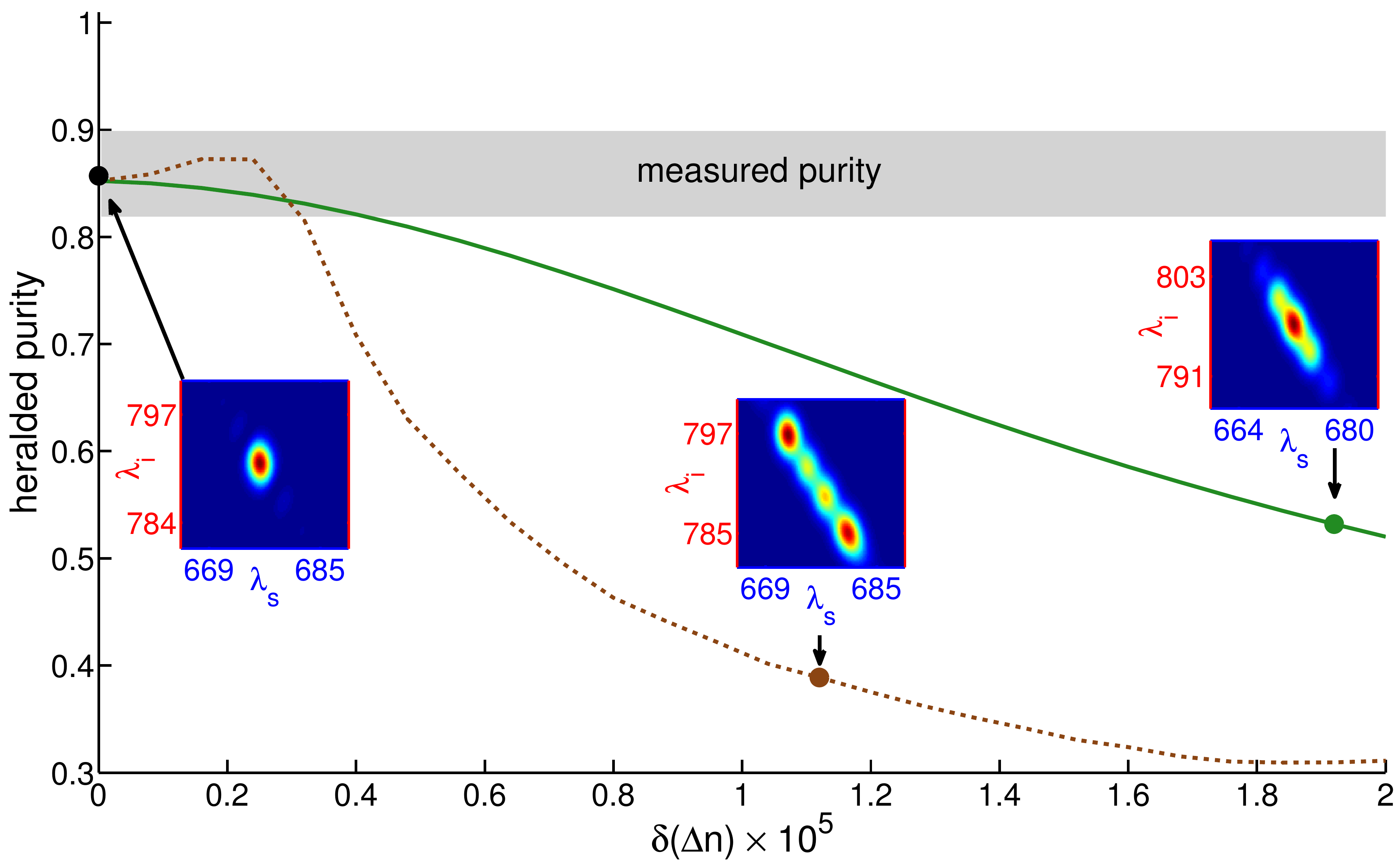}	
\caption{The heralded state purity resulting from two models of birefringence variation are shown.  Gradual variation (solid line) is approximated by a linear dependence $\Delta n(z) = \Delta n_0+\delta(\Delta n)\times z/L$. Rapid variation (dashed) is modelled by randomly fluctuating birefringence of mean $\langle\Delta n\rangle = \Delta n_0$ and standard deviation $\delta(\Delta n)$.  Random inhomogeneity increases the heralded purity for small $\delta(\Delta n)$ due to suppression of the peripheral lobes in the joint spectral amplitude.  All results assume L $= 4$ cm and $\Delta n_0=10^{-4}$.  Insets show representative joint spectral intensities.  The gray shaded region indicates the $95\%$ confidence interval of our measured heralded purity.}
\label{fig:inhomBiref}
\end{figure}

We consider two models of birefringence inhomogeneity and their effect on heralded state purity in Fig.~\ref{fig:inhomBiref}.  The effects of gradual changes during fabrication, for example variation in the writing laser power or local environment, are modeled as a birefringence that varies linearly along the length of the guide.  Rapid fluctuations, on the other hand, are described as a random variation in the birefringence of a specified mean and standard deviation.  For both models, the resulting phase-matching function is found by numerically integrating $\phi(\omega_s,\omega_i)=\int_0^L e^{iz\Delta k(z)}dz$, where $\Delta k(z)$ has an explicit spatial dependence.  The corresponding joint spectral amplitude in Eq.~\ref{eq:Hamiltonian} is then used to determine both the heralded purity via the Schmidt decomposition and the joint spectral intensities in the insets of Fig.~\ref{fig:inhomBiref}.  These simple models suggest our measured purity, to two standard deviations (gray band, Fig.~\ref{fig:inhomBiref}), corresponds to a birefringence inhomogeneity of $\delta_{max}(\Delta n)\leq 3\cdot 10^{-6}$. 

Undesirable birefringence variations in sources can reduce the quality of quantum interference.  Hong-Ou-Mandel interference of two single photons from identical sources produces a visibility equal to the photon purities.  In the ideal case of $\delta(\Delta n) = 0$ (left inset of Fig.~\ref{fig:inhomBiref}), we calculated a heralded purity of $0.98$ when using a $4.5$ nm filter on the heralding signal photon arm which still transmits $90\%$ of photons.  For an inhomogeneity $\delta(\Delta n)=3\cdot10^{-6}$, the heralded purity, and thus interference visibility, remains high at $0.95$ while the filter transmission is now $84\%$.  Accounting for source inhomogeneity, current fabrication methods thus appear sufficient to obtain high visibility interference from heralded sources.  

\section{Conclusion}
We have demonstrated an on-chip source of heralded single photons that achieves extremely low loss ($\eta_P = 80\%$) and high output state purity ($P=0.86$) without any single-photon filtering.  To our knowledge, no previous on-chip source has simultaneously demonstrated such low loss heralding of high purity states.  Achieving this performance relied on spectrally factorable photon emission which was enabled by the mode control allowed by source integration.  An estimate of birefringence inhomogeneity suggests current fabrication methods are sufficient for high visibility interference between multiple sources on the same chip.

Our source meets several loss thresholds for quantum-enhanced applications.  Interferometric phase estimation, using single-photon sources with $\eta_P=80\%$, can achieve a precision better than any classical probe field with the same number of photons~\cite{Datta2011}.  For linear optics quantum computing, the high $\eta_P$ and low $g^{(2)}_H(0)$ demonstrated here enables entangling gates to violate Bell inequalities without postselection~\cite{Jennewein2011}.  Exploiting this performance in future work is facilitated by the silica-chip architecture shared by our source and many recent integrated quantum optics experiments~\cite{Spring2013,Metcalf2013,Marshall2009,Shadbolt2011,Crespi2011,Owens2011,Crespi2013}.

In the longer term, multiplexing of sources either spatially with a switching network~\cite{Migdall2002,Shapiro2007} or temporally with quantum memories~\cite{Nunn2012} may provide a route to construct many-photon quantum states~\cite{Jennewein2011,Christ2012}.  Even with multiplexing, $\eta_P$ and $P$ still bound the composite source performance.  Therefore, optimizing individual source performance remains critical.

Our choice of SFWM in silica was motivated by the desire to minimize source loss.  One direction for future work is to build similar SFWM sources at telecom wavelengths, where silica loss is even lower.  As we have shown, the spectral flexibility of SFWM allows factorable states to be generated at these wavelengths with proper adjustment of $\Delta\lambda_p$ .  Quantum optics experiments in this spectral region, supplied by silica SFWM sources, could investigate a variety of fundamental scientific questions that can feasibly be tested with larger numbers of single photons.

\section*{Acknowledgements}
We thank Brian Smith, Christoph S\"{o}ller, and Josh Nunn for valuable insights. This work was supported by the UK EPSRC (EP/H049037/1, EP/E055818/1 and EP/H03031X/1), the EC project Q-ESSENCE (248095), the Royal Society, and the AFOSR EOARD. XMJ and NKL are supported by EU Marie-Curie Fellowships (PIIF-GA-2011-300820 and PIEF-GA-2010-275103). JBS acknowledges support from the United States Air Force Institute of Technology. The views expressed in this article are those of the authors and do not reflect the official policy or position of the United States Air Force, Department of Defense, or the U.S. Government.
\\ \\
\textit{Note: After preparing our manuscript, we learned about related advances in heralded single photon sources by SPDC in a pp-KTP waveguide~\cite{Harder2013}.}
\end{document}